\documentclass{elsart}

\usepackage[english]{babel}
\usepackage[dvips]{epsfig}

\begin{document}

\begin{frontmatter}

\title{Oscillating cosmological solutions within gauge theories of gravity}

\author{G.Vereshchagin}
\ead{griver@inbox.ru}

\address{Department of Theoretical Physics, Belorussian State University, F.Skoriny ave.4, 220050, Minsk, Republic of Belarus \\ International Center for Relativistic Astrophysics, Rome-Pescara, Italy}

\begin{abstract}
New type of nonsingular oscillating solutions for the Universe described by cosmological equations of gauge theories of gravity is presented. Advantages of these solutions with respect to existing nonsingular solutions within framework of general relativity and gauge gravity are discussed. It is shown in particular that these solutions have nonzero measure and stable on contraction stage unlike usual nonsingular solutions.
\end{abstract}


\end{frontmatter}

\section{Introduction}

In conventional cosmological scenario the Universe has a beginning in time, so called Big Bang, associated with initial singularity of Friedmann equations \cite{Kol90}. Singular state with infinite values of energy density and curvature is unacceptable from physical point of view and possibly marks the edge of applicability of General Relativity (GR). In spite of some progress in last decades, attempts to resolve this problem within GR had no success \cite{Linde},\cite{BMS}.

Correct description of the early Universe when the energy density was very high should be given by particle physics \cite{Linde}. Unfortunately experimental and observational data are still insufficient to discriminate between different models of elementary particles physics. However in most models scalar fields play an important role. One of the main discovery in modern cosmology is a possibility to explain a number of puzzles of Big Bang model by inclusion of short stage of exponential expansion of the Universe, so called inflation \cite{Linde},\cite{Olive}. During this stage the energy density of the Universe is dominated by scalar field (or fields) and transition to usual hot Universe is achieved by creation of relativistic particles at the end of inflation \cite{Reheating}.

Unfortunately even in inflationary models there is little chance to avoid cosmological singularity. The evolution of the Universe is usually considered from the planckian moment when description in terms of classical theory of gravitation becomes possible. At contraction stage the singular state is unavoidable in most models \cite{BMS},\cite{BGZH}. Some nonsingular solutions exist \cite{BGZH}-\cite{Mink03a} but they are not of great interest since their measure is equal to zero and they cannot be used to construct viable nonsingular cosmological models.

Alternative way to resolve problems of Big Bang model could be so called oscillating model \cite{Durrer} where the Universe never reaches singularity. In the recent literature such models are debated in connection in particular to ekpyrotic \cite{Ekp} and cyclic Universe models \cite{Cyc}.

From the other hand, deviations from Riemannian geometry can occur at very high energy density. In the case of more general spacetime GR should be replaced by gauge theories of gravity (GTG) that are natural generalizations of GR \cite{Hehl}. It was shown that nonsingular cosmological solutions within GTG exist in the case of spinning matter, usual matter \cite{Acta98} and matter in the form of scalar field \cite{mv1}. Moreover,  when energy density drops enough the deviation from GR becomes negligible and at late stages the Universe can be described by Friedmann equations that possess to generalize standard cosmology and resolve the problem of initial singularity.

In this paper we present new type of nonsingular cosmological solutions for the Universe filled by nonlinear scalar field within gauge theories of gravity. We show that our solutions have some advantages with respect to previously known nonsingular solutions. The paper is organized as follows. In section 2 cosmological equations of GTG are given. In section 3 oscillating nonsingular solutions are presented. In section 4 stability of our solutions is discussed. In section 5 discussion is given. The last section contains conclusions.

\section{Cosmological equations}

Homogeneous isotropic models in GTG are described by generalized Friedmann cosmological equations \cite{phlett80} that were obtained first within framework of Poincare gauge theory of gravity. Same equations are valid in metric-affine gravity. They read

\begin{equation}
\label{gcfe}
\frac{k}{R^2}+\left\{\frac{d}{dt}\ln\left[R\sqrt{\left|1-\beta\left(\rho-
3p\right)\right|}\,\right]\right\}^2=\frac{8\pi }{3M_{p}^{2}}\,\frac{\rho-
\frac{\beta}{4}\left(\rho-3p\right)^2}{1-\beta\left(\rho-3p\right)}
\, ,
\end{equation}

\begin{equation}
\label{Heqn1}
\frac{\left[\dot{R}+R\left(\ln\sqrt{\left|1-\beta\left(\rho-
3p\right)\right|}\,\right)^{\cdot}\right]^\cdot}{R}= -\frac{8\pi
}{6M_{p}^{2}}\,\frac{\rho+3p+\frac{\beta}{2}\left(\rho-3p\right)^2}{
1-\beta\left(\rho-3p\right)}\, .
\end{equation}
where $R(t)$ is the scale factor in the Robertson-Walker metric,
$k=-1,0,+1$ for open, flat and closed models respectively, $M_{p}$ is
planckian mass, $\rho(t)$ and $p(t)$ are energy density and pressure, $\beta$ is indefinite parameter, and a dot denotes the differentiation with respect to time\footnote{Hereinafter the system of units with $\hbar=c=1$ is used.}.

These equations contain the only indefinite parameter $\beta$ that is a combination of gravitational lagrangian coefficients, having a dimension of inverse energy density. Solutions of cosmological equations have essentially non-einsteinian behavior when the energy density exceeds the critical value $\beta^{-1}$. Experimental constraints on parameter $\beta$ are very poor and one can say only that $\beta^{-1}\leq M_p^4$.

Besides equations (\ref{gcfe}-\ref{Heqn1}) gravitational equations of GTG lead to the following relation for torsion function $S$ and nonmetricity function
$Q$

\begin{equation}
\label{SQ}
S-\frac{1}{4}Q=-\frac{1}{4}\frac{d}{dt}\ln[1-\beta(\rho-3p)].
\end{equation}
In Poincare gauge theory of gravity $Q=0$ and equation (\ref{SQ}) determines the torsion function.  In metric-affine theory of gravity equation (\ref{SQ}) may describe three kinds of models: in the Riemann-Cartan spacetime $Q=0$, in the Weyl spacetime $S=0$, in the Weyl-Cartan spacetime ($S\neq 0, Q\neq 0$), the function $S$ is proportional to the function $Q$ \cite{garCQG}.

The conservation law has the usual form

\begin{equation}
\label{cl}
\dot{\rho}+3H(\rho+p)=0,
\,\,\,\,\,\,\,\,\,\,\,\,\,\,\,\,\,\,\,\,\,\,\,
H(t)\equiv\frac{\dot{R}}{R},
\end{equation}
where $H(t)$ is a Hubble parameter.

In the case of nonlinear scalar field minimally coupled to gravity the energy density and pressure are

\begin{equation}
\begin{array}{rcl}
\label{phicon}
\rho=\frac{1}{2}\dot{\varphi}^2(t)+V(\varphi),
\,\,\,\,\,\,\,\,\,\,\,\,\,\,\,\,\,\,\,\,\,\,\,
p=\frac{1}{2}\dot{\varphi}^2(t)-V(\varphi),
\end{array}
\end{equation}
where $V(\varphi)$ is effective potential of the scalar field $\varphi(t)$.

Then taking into account (\ref{cl}-\ref{phicon}) the cosmological equations (\ref{gcfe}-\ref{Heqn1}) can be rewritten in the following way \cite{Mink02},\cite{mv1},\cite{Ver03}

\begin{equation}
\label{gcfe2}
\begin{array}{l}
\begin{displaystyle}
\frac{k}{R^2}+\left[H\left(1-3\beta\dot{\varphi}^2 Z^{-1}\right)-
3\beta V' \dot{\varphi} Z^{-1}\right]^2=
\end{displaystyle} \\
\begin{displaystyle}
\;\;\;\;\;\;\;\;\;\;\;\;\;\;\;\;\;\;\;\;
\;\;\;\;\;\;\;\;\;\;\;\;\;\;\;\;\;\;\;\;
\frac{8\pi }{3M_{p}^{2}}\,
\left[\frac{1}{2}\dot{\varphi}^2+V-\frac{\beta}{4}\left(4V-
\dot{\varphi}^2\right)^2\right]\,Z^{-1},
\end{displaystyle}
\end{array}
\end{equation}

\vskip4mm

\begin{equation}
\label{Heqn2}
\begin{array}{l}
\begin{displaystyle}
\dot{H}\left(1-3 \beta\dot{\varphi}^2
Z^{-1}\right)+H^2\left(1+15\beta\dot{\varphi}^2 Z^{-1}-18\beta^2\dot{\varphi}^4
Z^{-2}\right)+
\end{displaystyle} \\
\begin{displaystyle}
\;\;\;\;\;\;\;\;\;\;\;\;\;
12\beta\dot{\varphi}V'Z^{-1}
\left(1-3\beta\dot{\varphi}^2Z^{-1}\right)H-
\end{displaystyle} \\
\begin{displaystyle}
\;\;\;\;\;\;\;\;\;\;\;\;\;\;\;\;\;\;\;\;\;\;\;\;\;\;\;\;\;
3\beta Z^{-1}\left(V''\dot{\varphi}^2-(V')^2+
6\beta\dot{\varphi}^2(V')^2Z^{-1}\right)=
\end{displaystyle} \\
\begin{displaystyle}
\;\;\;\;\;\;\;\;\;\;\;\;\;\;\;\;\;\;\;\;
\;\;\;\;\;\;\;\;\;\;\;\;\;\;\;\;\;\;\;\;
\frac{8\pi}{3M_{p}^{2}}\left[V-\dot{\varphi}^2-\frac{\beta}{4}\left(4V
-\dot{\varphi}^2\right)^2\right]\,Z^{-1},
\end{displaystyle}
\end{array}
\end{equation}

\vskip4mm

where $Z(t)=1+\beta\left[\dot{\varphi}^2(t)-4V(\varphi(t))\right]$,
$V'=\frac{dV(\varphi)}{d\varphi}$, $V''=\frac{d^2V(\varphi)}{d\varphi^2}$.

The conservation law (\ref{cl}) in the case under consideration is reduced to the scalar field equation
\begin{equation}
\label{scl}
\begin{array}{rcl}
\ddot{\varphi}+3H\dot \varphi=-V'.
\end{array}
\end{equation}

\section{Oscillating solutions}

Numerical investigation of nonsingular solutions of equations (\ref{gcfe2}-\ref{scl}) was carried out in \cite{Mink02},\cite{mv1},\cite{Ver03},\cite{Mink03b}. It was shown in particular that it is possible to obtain nonsingular solutions if parameter $\beta$ is negative and potential energy of the scalar field dominates its kinetic energy at the bounce. In the following we shall assume simple effective potential usually considered in inflationary models \cite{Linde}

\begin{equation}
\label{fi4}
V(\varphi)=\frac{1}{4}\lambda \varphi^4,
\end{equation}
where $\lambda$ the a coupling constant.

The constant $\lambda$ must satisfy the condition $\lambda\sim10^{-12}$ \cite{Linde} to obtain density fluctuations after inflation as small as $\frac{\delta\rho}{\rho}\sim10^{-4}$.

\begin{figure}[ht]
\begin{center}
\epsfxsize=5in
\epsfbox{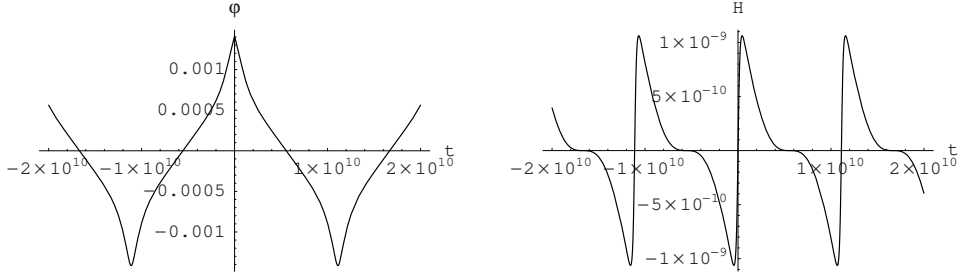}
\caption{Oscillating solution with negative parameter $\beta$, corresponding to closed cosmological model.}
\label{fig:4}
\end{center}
\end{figure}

The typical solution in the case of closed model is shown at fig.\ref{fig:4}. Initial conditions for this solution are $\lambda=10^{-12},\;\varphi_0=\sqrt2\cdot10^{-3}\,M_p,\; \dot{\varphi}_0=0,\; \beta=-10^{25}\,M_p^{-4}.$ The duration of expansion is $5\cdot10^9\,M_p^{-1}$.

Initially the scalar field is placed at some distance from the value $\varphi=0$ and the kinetic part of the energy density is zero. Therefore at the beginning the effective equation of state is $p\approx-\rho$ that is one of requirements for inflation. At the same time the Hubble parameter is zero and its derivative is positive that means transition from contraction to expansion. In course of evolution the scalar field goes to zero; potential energy is transformed into kinetic one. The Universe expands.

When the scalar field reaches the minimum of effective potential $V(\varphi)$ its energy is equal to the kinetic energy. The expansion is replaced by contraction in that moment. Then the scalar field slows down and the energy density is transferred back to the effective potential and we arrive to the initial state. The function (\ref{SQ}) oscillates with the same frequency and sign as Hubble parameter. 

Oscillations of the scalar field are usual for inflationary models in the framework of GR, but the stage of oscillations has nothing to do with inflation itself. It corresponds to the stage of reheating after inflation during which usual particles are created from the scalar field \cite{Linde}. In our case however oscillations in scalar field occur together with oscillations of the whole Universe.

Note, that in the model under consideration, condition of inflation $\dot H\ll H^2$ is satisfied during expansion. The duration of the phase of expansion, however, is close to the inverse Hubble parameter, i.e. to the age of the Universe. Thus, the expansion rate is very small to allow one to speak about the real inflation.

It is clear, that usual mechanism of reheating cannot work in this model because it was developed for monotonically varying background. However, recently a very efficient mechanism of preheating was found \cite{IP}, when the only one oscillation is necessary for particles to be created. If some mechanism produce scalar field at the stage of contraction, the solution will describe the regular cosmological model in which very rapid transition from contraction to expansion appears when the energy density of the Universe is concentrated in the scalar field. The rest of the time the Universe is usual Friedmann one.

It is necessary to note that particular form of the effective potential (\ref{fi4}) is not so important. As numerical calculations show oscillating solutions are quite generic feature for GTG. The only restrictions are that there must be a local minimum of the effective potential and it should be sufficiently flat.

\section{Stability with respect to initial conditions}

In this section we show using intuitive approach that our solutions have one advantage with respect to known solutions obtained in GR as well as GTG, namely our solutions are stable on the stage of contraction.

\begin{figure}[ht]
\begin{center}
\epsfxsize=5in
\epsfbox{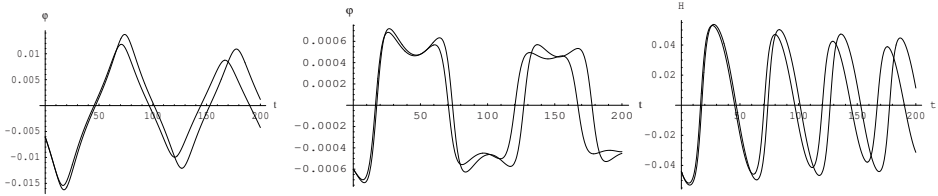}
\caption{Cosmological parameters in oscillating solutions with different initial conditions. Both solutions are bounded that is an indication of stability of oscillating solutions.}
\label{fig:b}
\end{center}
\end{figure}

With this goal we vary initial conditions, namely values of $\varphi_0$, $\dot\varphi_0$ and $H_0$ at the beginning (starting at some moment on contraction stage) and see what happens with solutions. Example is given at fig.\ref{fig:b} for finite time intervals\footnote{It can be seen that the amplitude of oscillations of cosmological parameters is decreasing in time. If one follows solutions for a longer time it can be found that the amlitude itself oscillates reminding modulations. The frequency of modulations is typically $10^3$ times smaller than the amplitude of oscillations.}. Both solutions are bounded, so singularity never appears. They are most sensitive to variation of initial value of the derivative $\dot\varphi_0$ but if condition $\frac{1}{2}\dot\varphi_0\ll V(\varphi_0)$ is satisfied solutions look very similar. Sensitivity to initial value of the Hubble parameter is very weak. Note that the frequency of oscillations depends on initial conditions even for small variations of $\varphi_0$.

In general case together with our oscillating solutions singular solutions are present. However ranges for initial conditions allowing bounded solutions are finite although quite narrow (at least in $\dot\varphi_0$ direction). These finite regions represent 'islands of stability' were nonsingular solutions exist.

Finally we plot 3-dimensional phase trajectories for our solutions with varying initial conditions. It is clear that phase trajectories lay at some 2-dimensional hypersurface (probably slice). Each trajectory is a closed curve that means that solutions are indeed oscillating. These closed curves do not remind circles and therefore oscillations are not harmonic and likely cannot be described by analytical functions.

\begin{figure}[ht]
\begin{center}
\epsfxsize=5in
\epsfbox{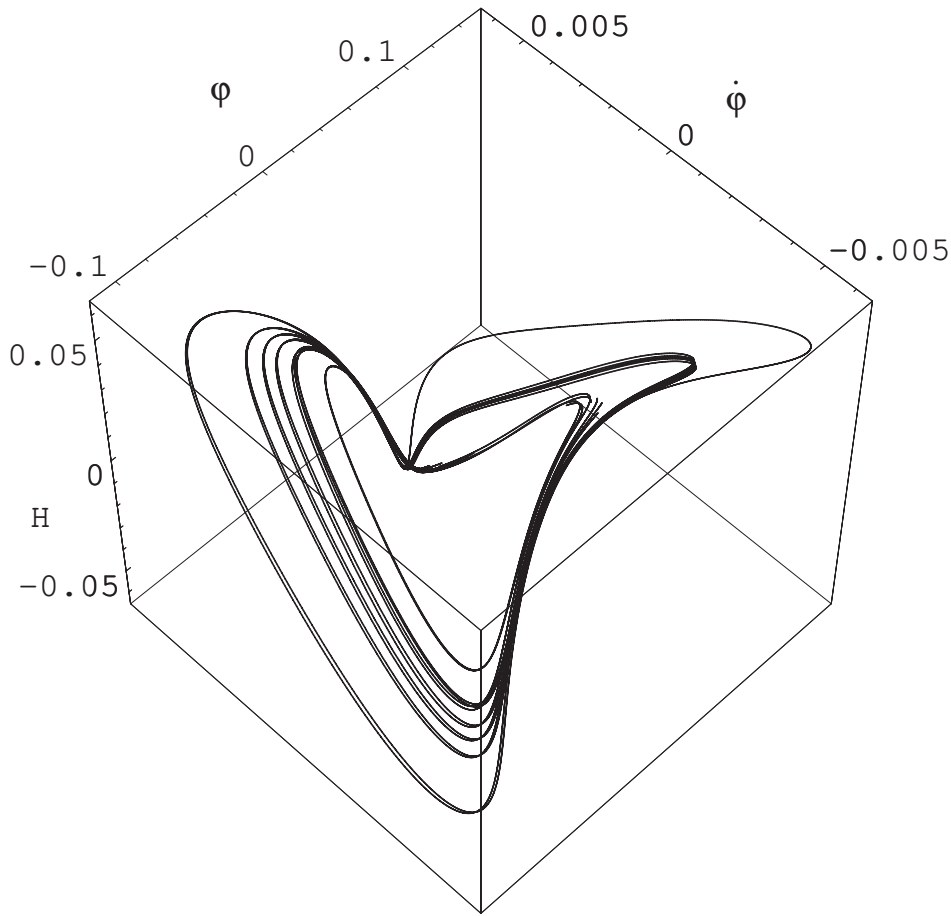}
\caption{3-dimensional phase portrait of oscilating solutions. They all lay at some 2-dimensional curved hypersurface (or slice).}
\label{fig:3d}
\end{center}
\end{figure}

\section{Discussion}

In this section we would like to carry out comparison to existing nonsingular solutions obtained previously within framework of GR as well as GTG for cosmological models with single scalar field.

Nonsingular cosmological solutions for the Universe filled by scalar field within GR are known since a long time \cite{Star} and are discussed in recent paper \cite{Mink02},\cite{Kan}. It was shown that such solutions exist in GR only for closed models. Unfortunately their measure is equal to zero and moreover on the stage of contraction they are unstable \cite{BGZH},\cite{KLM}. For most solutions in course of contraction effective equation of state becomes $p=\rho$ and this leads to singularity. Therefore such solutions are of certain interest in connection with the initial singularity problem but they do not cure for final singularity.

Moreover, on the stage of expansion there is transition from vacuum dominated expansion regime to Friedmann expansion due to particle creation process. It is well known that such process is irreversible \cite{KLM}. Appearing and subsequent domination of the scalar field on contraction stage is an obscure question.

Nonsingular inflationary solutions with nonlinear scalar field were obtained within framework of GTG first in \cite{mv1} in the case of initial conditions used in chaotic inflation \cite{Lin83}. In subsequent papers \cite{Mink02},\cite{Mink03b} such solutions were discussed to some extent. The crucial property of such solutions is presence of inflationary stage during expansion; at the same time quasi-de-Sitter contraction stage is present in all of them. 

It is necessary to stress that similar solutions exist within the framework of GR \cite{Mink03a} most of them are singular on contraction stage since equation of state $p=\rho$ is much more probable. Careful investigation of this problem for the case of flat cosmological models with massive scalar field in GTG \cite{Ver03} lead to similar conclusions\footnote{However effective equation of state is $p=\frac{\rho}{3}$ in this case.}. One can perform simple numerical analysis similar to ours and change values of $\varphi_0$, $\dot\varphi_0$ or $H_0$ on contraction stage. It can be shown that even little changes lead to unbounded solutions. This is a typical property of nonlinear equations. Therefore in our opinion oscillating solutions have important advantage although there is no inflation on expansion stage.

Finally it should be stated that our solutions can be obtained in the wide regions of initial conditions. Moreover, since maximal value of energy density depends on parameter $\beta$, oscillating solutions can be obtained for the case when $\rho\ll M_p^4$, the Universe can never enter epoch of quantum gravity and can be described by classical theory of gravitation.

\section{Conclusions}

In this paper we give a new type of cosmological solutions of GTG for the Universe filled by nonlinear scalar field. Numerical analysis shows that these solutions are stable on contraction stage unlike solutions with inflationary stage. If any mechanism of reheating can work in our case, it can be possible to build nonsingular cosmological model on the basis of our solutions.

\end{document}